\documentstyle{elsart}
\begin{document}
\begin{frontmatter}
\title{Giant Cyclones in Gaseous Discs \\
of Spiral Galaxies}

\author[Inasan]{A.\,M.\,Fridman, O.\,V.\,Khoruzhii, E.\,V.\,Polyachenko}
\author[GAISH]{A.\,V.\,Zasov, O.\,K.\,Sil'chenko}
\author[SAO]{V.\,L.\,Afanas'ev, S.\,N.\,Dodonov, A.\,V.\,Moiseev}
\address[Inasan]{Institute of Astronomy, 48 Pyatnitskaya Str., Moscow,
109017 Russia}
\address[GAISH]{Sternberg Astronomical Institute, Universitetskii pr.
13, Moscow, 119899 Russia}
\address[SAO]{Special Astrophysical Observatory, Nizhnyi Arkhyz,
Karachaevo-Cherkessia, 357147 Russia}

\begin{abstract}
We report detection of giant cyclonic vortices in the gaseous
disc of the spiral galaxy NGC 3631 in the reference frame rotating with
the spiral pattern. A presence of such structures was predicted by the
authors for galaxies, where the radial gradient of the perturbed velocity exceeds
that of the rotational velocity. This situation really takes place in NGC
3631.
\end{abstract}
\end{frontmatter}

\section{Introduction}
In  previous papers \cite{1},\,\cite{2} authors have shown that besides
the spiral arms, well known for more than one and a half century in
spiral galaxies as structures in the brightness distribution, there also
exist structures, revealing themselves in the velocity field. The latter
structures
have the appearance of giant anticyclonic vortices: their rotation is
opposite to the rotation of the galactic disc.
The centers of these vortices are located near the zero points of the force field
(Lagrange points $L_4$ and $L_5$, see \cite{3}). In spiral galaxies these
points are fixed in the close vicinity of the corotation circle, where the
velocity of the rigid-body rotating spiral pattern coincides with the velocity
of the differentially rotating disc. Being in such a position, the centers of
the anticyclones turn out to be fixed with respect to the spiral
arms and stationary if the spiral structure is stationary. If the spiral structure
develops due to some instability in the disc, then the arms and vortices
arise simultaneously and grow with the same growth rate \cite{4}. As a
result a unified spiral--vortex structure forms.

The growth of the perturbation amplitude leads rather often to
the saturation of the instability, after which a spiral--vortex structure
becomes stationary. From the very beginning this scenario implies
that spiral arms in discs of galaxies have a wave
nature, that satisfies modern theoretical conceptions \cite{5} and
agrees well with the results of the velocity field analysis in spiral
galaxies (\cite{2},\,\cite{6}).

A typical rotation velocity field of a galactic disc is
represented schematically in Fig.~1a. There is no way to close the
trajectories of particles in the laboratory (inertial) reference frame
by any radial component of the velocity (arrows radially directed
in Fig.~1a). The situation is opposite in the rotating reference frame,
where anticyclonic motion takes place under condition of
appropriate azimuthal variation of the radial component (see Fig.~1b).
Note that in the latter case it is not necessary to have large radial
velocities to form a vortex near the circle corotating with
the reference frame due to small values of the azimuthal velocities
in this region.

It is not accidental that the radial velocity was chosen as a periodic
function of the azimuth:  in the spiral density wave such periodicity
should take place both for the radial $\tilde V_r$ and for the azimuthal $\tilde
V_\phi $ perturbed velocities.  Arrows for $\tilde {V_\phi} $ are not
shown in Fig.1.  Taking into account the perturbed azimuthal velocity
will not change the picture qualitatively, if
\begin{equation}
\label{1a}
|\partial\tilde{ V_\phi}/ \partial r| < |d V_{circ} / dr|,
\end{equation}
where $V_{circ}$ $\equiv$
$V_{rot}-\Omega_{rf}\,r$ is the circular velocity in the reference frame
rotating with the angular velocity $\Omega_{rf}$. Note that
even a weak inequality is quite enough to form only anticyclonic vortices
like in Fig.\,1b (for more details see \cite{6aa}). Under the
opposite condition:
\begin{equation}
\label{1b}
|\partial\tilde {V_\phi}/\partial r| > |d V_{circ}/dr|
\end{equation}
not only anticyclonic, but also cyclonic vortices can appear \cite{6a}.
In the next section we will consider both these cases in a schematic model
of a galactic disc with a well-defined two-armed spiral structure
(Grand-Design galaxy).

\section{Qualitative description of anticyclone and cyclone formation in
 a disc of Grand-Design galaxy}

Let us assume that two-armed Grand-Design spiral structure exists in a
differentially rotating gaseous disc. Then the functions describing the
perturbed surface density and velocity components, related to the density
wave, may be approximated in the form:
\begin{equation}
\label{1}
\tilde{\sigma}(r,\,\varphi ) = C_\sigma(r)\,\cos [2\varphi- F_\sigma (r) ]
\, ,
\end{equation}
\begin{equation}
\label{2}
\tilde V_r(r,\,\varphi ) = C_r(r) \, \cos [ 2\varphi- F_r (r) ] \, ,
\end{equation}
\begin{equation}
\label{3}
\tilde V_\varphi (r,\,\varphi ) = C_\varphi(r) \, \cos  [ 2\varphi-
F_\varphi (r) ] \, .
\end{equation}

As we consider the momentary picture of the spiral structure, a time
dependence is absent in the equations above.

It is natural to ask, whether it is possible to use expressions
(\ref{1})--(\ref{3}) for the description of the dynamics of spiral density
waves with a finite amplitude. We have checked it in
\cite{6},\,\cite{9} by the example of the galaxy NGC~3631, where cyclones
have been found just now (see Section III below). The operating
sequence of the check was the following.  First, we analysed the azimuthal
expansion in the harmonic series of the observed surface brightness of
this galaxy at different galactocentric radii. A domination of the second
Fourier harmonic over others (see Fig.~1 in \cite{9}) proved the
correctness of the approximation expressed in Eq.(\ref{1}).

Then the
observed line-of-sight velocity field $V^{obs}(r, \varphi)$ was analysed.
This velocity can be expressed as (see \cite{2},\,\cite{7}):
\begin{equation}
\label{4}
V^{obs} (r, \varphi )\,=\, V_s \,+\,
V_\varphi(r,\varphi ) \, \cos \varphi \,+\, V_r (r,\varphi) \, \sin
\varphi \, \sin i \,+\, V_z (r,\varphi) \,\cos i \,.
\end{equation}
Here $i$ is an inclination angle (angle between the rotation axis of
galaxy and line of sight) and $V_s$ is systemic velocity of a galaxy
(velocity of the center of mass of the galaxy).

Within the frame of the present model
\[
V_r(r,\varphi) ~=~ \tilde V_r (r,\varphi)\,,
\]
\begin{equation}
\label{5}
V_\varphi(r,\varphi) ~=~ V_{rot}(r) ~+~ \tilde V_\varphi (r,\varphi),
\end{equation}
\[
V_z(r,\varphi) ~=~ \tilde V_z (r,\varphi) ~=~
C_z(r)\cos[2\varphi ~-~ F_z(r)] \, ,
\]
where $\tilde V_r$, $\tilde V_\varphi$ are determined by (\ref{2}) and
(\ref{3}). Substituting (\ref{5}) in (\ref{4}) one obtains the model
representation of the Fourier expansion for $V^{obs}(r,\varphi)$.

Equating the coefficients of the model expansion to the coefficients
calculated from observational data,
\begin{equation}
V^{obs}(r, \varphi) ~=~ V_s ~+~ \sum\limits_n \left(a_n^{obs}\, \cos
n\varphi  + b_n^{obs}\, \sin n\varphi \right) \,,
\label{6s}
\end{equation}
we obtain the following set of relations between unknown characteristics of
the vector velocity field of the galaxy and the Fourier coefficients of the observed
line-of-sight velocity field:
\begin{equation}
V_{rot} ~+~ \frac12(C_r \sin F_r + C_\varphi
\cos F_\varphi) ~=~ a_1^{obs}, \label{6} \end{equation} \begin{equation}
-C_r \cos F_r + C_\varphi \sin F_\varphi = 2b_1^{obs}, \label{7}
\end{equation} \begin{equation} C_z \cos F_z \cos i ~=~ a_2^{obs},
\label{8}
\end{equation}
\begin{equation}
C_z \sin F_z \cos i ~=~ b_2^{obs},
\label{9}
\end{equation}
\begin{equation}
-C_r\sin F_r ~+~ C_\varphi \cos F_\varphi ~=~ 2a_3^{obs},
\label{10}
\end{equation}
\begin{equation}
C_r\cos F_r ~+~ C_\varphi \sin F_\varphi ~=~ 2b_3^{obs}.
\label{11}
\end{equation}

From (\ref{6})-(\ref{11}) we see that the perturbed velocity components
$\tilde{ V_r,} \tilde{ V_\varphi}$ and $\tilde {V_z}$, written in the form
of (\ref{2}), (\ref{3}) and (\ref{5}) give contributions only in the
first three harmonics of the Fourier expansion. This is really observed in
the case of NGC 3631 (see Fig.~3 in \cite{6}), where the first three
Fourier harmonics of the line-of-sight velocity field dominate over the
higher-order harmonics. This gives evidence for the correctness of
the representation of the perturbed velocity components in the form (\ref{2}),
(\ref{3}) and (\ref{5}) for the galaxy NGC~3631.

To draw schematically the spiral--vortex structure in different
reference frames we need relations
between the phases   $F_r(r),~ F_\varphi (r)$ and $
F_\sigma (r)$. These relations were derived in \cite{7} from the set of linearized
equations. In spite of this simplification, an excellent
qualitative agreement was found between the phase of the ''modified''
third harmonic (see below) of the observed line-of-sight velocity field and $F_\sigma
(r)$ (see Fig. 5 in \cite{6}), which enables to use the phase relations
obtained in \cite{7} for a qualitative description of the residual
velocities. The modified third harmonic is a two-armed spiral with phase
$F_3 - \pi/2,$ i.e. $\propto \cos (2 \varphi - F_3 + \pi/2),$ where
$F_3$ is the phase of the original radial-velocity-field third harmonic.

Below a list of these relations is presented.

a) Relations between the phases of the {\it radial} velocity
$\tilde V_r$ and the surface density perturbation $\tilde
\sigma$.

\begin{itemize}
\item On the corotation circle $ r~=~r_c$ :
\begin{equation}
F_r~=~F_\sigma~-~\displaystyle\frac\pi{2},
\label{12}
\end{equation}
whence
\begin{equation}
\tilde V_r (r_c, \varphi) ~=~-C_r(r_c)~\sin[2\varphi~-~F_\sigma(r_c)],
\label{13}
\end{equation}
i.e. on the corotation circle the extrema of $\tilde V_r$ coincide with
the zeroes of $\tilde\sigma$ and vice versa.

\item Inside the corotation circle, $r < r_c$,
\begin{equation}
F_r ~=~ F_\sigma~+~\pi,
\label{14}
\end{equation}
whence
\begin{equation}
\tilde V_r (r, \varphi) ~=~ -C_r (r) \cos[2\varphi - F_\sigma(r)],\quad
r<r_c,
\label{15}
\end{equation}
i.e. inside the corotation circle the maxima of
$\tilde V_r$ coincide with the minima of $\tilde\sigma$ and vice versa.

\item Outside the corotation circle, $r>r_c$,
\begin{equation}
F_r = F_\sigma,
\label{16aa}
\end{equation}
whence
\begin{equation}
\tilde V_r (r, \varphi) ~=~ C_r (r) \cos[2\varphi - F_\sigma(r)],\quad
r>r_c,
\label{16}
\end{equation}
i.e. outside the corotation circle the maxima and minima of $\tilde V_r$
coincide, respectively, with the maxima and minima of $\tilde\sigma$.
\end{itemize}

b) Relation between the phases of the {\it azimuthal} velocity
$\tilde V_\varphi $ and the surface density $\tilde\sigma$.

\begin{itemize}
\item
In all cases, both on the corotation circle, $r=r_c$, and out of it,
$r\not= r_c$, we have
\begin{equation}
F_\varphi~=~F_\sigma~+~\displaystyle\frac\pi{2},
\label{17}
\end{equation}
therefore
\begin{equation}
\tilde V_\varphi  (r, \varphi) ~=~ C_\varphi (r) \sin[2\varphi -
F_\sigma(r)],
\label{18}
\end{equation}
i.e. the extrema of $\tilde V_\varphi $ coincide with the zeroes of
$\tilde\sigma$ and vice versa throughout the disc.
\end{itemize}

Using the expressions (\ref{13}), (\ref{15}), (\ref{16}), and (\ref{18}) we
can draw a scheme for the velocity field with vortices overlayed on the
scheme of the surface density distribution (Fig.~2).  In Fig.~2b we
can see the presence of cyclones and anticyclones in
the field of residual velocity (that corresponds a
reference frame locally corotating with the disc at each radius,
$V_{circ}(r) = 0$). This fact by itself does not depend on the amplitude
of the spiral density wave, although the sizes of the vortices depend on
this amplitude.

Lindblad and Langebartel \cite{8} were the first to calculate the
field of displacements of the stars in the gravitational potential of a bar.
The form of the star displacement field resembles the system of two
cyclones  and two anticyclones, which is similar our schematic Fig.~2b.
Refering on \cite{10} Lynden- Bell \cite{11} has noted the existence of
cyclonic and anticyclonic trajectories in Fig.~9 of\cite{8}.
A similar picture of four vortices in the velocity field of a gaseous disc
-- two cyclones on the bar and two anticyclones between the spiral arms --
was obtained in a recent paper \cite{cont}.

In the reference frame rotating with the angular velocity
$\Omega_{ph}$ of the two-armed spiral pattern the picture of vortices
differs qualitatively from the previous one. In the case (1) when the
gradient of the perturbed azimuthal velocity is less than the circular
velocity gradient, i.e.  the anticyclonic shear dominates over the
cyclonic one, the appearance of cyclones is impossible -- we can see only
two anticyclones with centers near the corotation circle between two
spiral arms (Fig. 2c). In the opposite case (2) when the
gradient of the perturbed azimuthal velocity dominates over the circular
velocity gradient, the location of the cyclones is determined by the place
of this domination.  Cyclones can survive almost in the same places as
in Fig.~2b, i.e. on the corotation circle, but their sizes will be
smaller.  There are also two other variants when the cyclone center moves
either inside or outside the corotation circle along the zero lines of
the radial velocity. Finally, both last these variants can be realized
simultaneously.  In this case four cyclones exist in the velocity
field as is shown in Fig.~2d.

\section{ Discovery of Giant Cyclones in the Gaseous Disc of the Spiral
Galaxy NGC~3631 with the 6m telescope in SAO}

Finding giant vortices in a spiral galaxy requires the reconstruction of
the vector velocity field in the reference frame
rotating with the angular velocity of the spiral pattern. Thus, first of
all the corotation radius should be determined. In this paper for NGC~3631
we use the position of the corotation derived in \cite{9} on the base of
the Fourier analysis of the observed line-of-sight velocity field. The
observations of NGC 3631 were carried out with the scanning Fabry-Perot
interferometer with Russian 6m telescope. The details of observations were
described in \cite{6}.

It is necessary to determine $\tilde V_r $ and $V_\varphi = V_{rot} +
\tilde V_\varphi$, i.e. to find five unknown functions:  $V_{rot}(r)$,
$C_r(r)$, $C_\varphi (r)$, $F_r(r)$, $F_\varphi (r)$ (see
Eqs.~(\ref{2}),\,(\ref{3}), and (\ref{5})).  These five functions are
connected with the characteristics of the observed velocity field by the four
relations (\ref{6}), (\ref{7}), (\ref{10}), (\ref{11}). An additional
condition required to close the system should have a theoretical origin.
Unfortunately, up to now a reliable condition valid for any density wave
amplitude is not available.  Several possibilities discussed in the
literature \cite{2},\,\cite{7} have limited applicability. To overcome
this difficulty, we propose the following approach.

Among the functions listed above, $V_{rot}(r)$ could be the most reliably
estimated on the base of independent observational data. For this purpose
the equilibrium condition of the gaseous disc rotating in
a gravitational potential $\Psi$ can be used
\begin{equation}
\displaystyle\frac{V^2_{rot}}r = \displaystyle\frac{\partial
\Psi}{\partial r}.
\label{19}
\end{equation}
The latter is determined from the mass
distribution in a galaxy or its surface brightness maps assuming a costant
mass-to-light ratio to be known. To calculate the right hand side of
the equation (\ref{19}) we use a three-component dynamical model of a spiral
galaxy similar to \cite{12}. In spite of the roughness of the model,
the $V_{rot}$ curve found by this method is in the region
between $(a_1)_{\min}$ and $(a_1)_{\max}$ (see Fig.~3;
a peculiarity of the thick curve is described below).

The same result is obtained by another way. From equation (\ref{6}) it
follows that the difference $|a_1-V_{rot}|$ cannot exceed the amplitudes
$C_r$ and $C_\varphi$, which, in turn, are connected by equations
(\ref{7}), (\ref{10}) and (\ref{11}) with the Fourier coefficients
$b_1^{obs}$, $a_3^{obs}$, and $b_3^{obs}$, determined from our
observations. In Fig.~4 the radial behaviour of $b_3^{obs}-b_1^{obs}$ $=$
$C_r \, \cos \,F_r$ and $b_3^{obs}+b_1^{obs}$ $=$ $C_\varphi \, \sin
\,F_\varphi$ is demonstrated. The extrema of these functions allow to
estimate the amplitudes $C_r$ and $C_\varphi$. According to
Fig.~4, one can conclude that in NGC~3631 a maximum value of the amplitude
of the residual velocities can reach at $60$~km/s, i.e.
\begin{equation}
|a_1 - V_{rot}|_{\max} \le 60 {\rm ~km/s}.  \label{20}
\end{equation}

The conditions (\ref{19}) and (\ref{20}) do not allow to calculate the
function $V_{rot}(r)$ exactly. Nevertheless, they set limits on the
variations of both the amplitude and the form of $V_{rot}(r)$. Within
these limits, we choose a set of trial curves (Fig.~3) and analyse the
velocity field obtained from Eqs.~(\ref{6})--(\ref{11}) for a given
$V_{rot}(r)$.

The general views of the restored velocity fields are qualitatively
similar for the full set of the rotation curves. In particular, in all
cases, except the most extreme ones, distinct areas of anticyclonic as
well as cyclonic shear flows exist. However, only rare examples demonstrate
the presence of areas with trapped fluid particles moving along stream
lines bounded by closed separatrices. Such behavior could be a natural
consequence of the quasi-stationary character of the density wave. It
is an argument in favour of choosing just these examples of
$V_{rot}(r)$ as the best-corresponding to the real dynamical model of
the galaxy. The best example of $U_{rot}$ from the point of view of the
stationarity of the stream lines is shown by the thick curve in Fig.3.
In other examples the vortices have not close separatrices and their
stream lines are spirals - strongly inwards or outwards.

In Fig.~5 we can see the amplitudes of radial ($\tilde V_r$) and
azimuthal ($\tilde V_\varphi$) components of the residual velocity. These
amplitudes, $C_r$ and $C_\varphi$ were obtained from four equations
(\ref{6}), (\ref{7}), (\ref{10}), and (\ref{11}) after substituting in
them $V_{rot}(r)$ in the form drawn in Fig.~3 by the thickest line.

Using the functions $C_r$ and $C_\varphi$ in Fig.5 let us show, that the
necessary condition (\ref{1b}) of the cyclone formation is satisfied in the
galaxy NGC 3631
outside and inside the corotation circle and is not satisfied on the
corotation circle. In the reference frame, rotating with the angular
velocity of the spiral pattern ($\Omega$ $=$ $\Omega_{ph}$), the inequality
(\ref{1b}) can be rewritten in the form
\begin{equation}
\left| \frac {\partial \tilde V_\varphi}{\partial r}\right| ~>~ \left|
\frac {d V_{circ}}{d r}\,  \right| ~\equiv~ \left| \frac {d }{d
r}\, (V_{rot} - \Omega_{ph}r)\right| \,.
\label{v1}
\end{equation}
Substituting here the expression (\ref{3}) for $\tilde V_\varphi$, we
obtain:
\begin{equation}
\left| C'_\varphi \, \cos(2 \varphi - F_\varphi) +F'_\varphi \, C_\varphi
\, \sin(2\varphi-F_\varphi)\right| ~>~ \left| V'_{circ}\right|\,,
\label{v2}
\end{equation}
where "prime" denotes the derivative with respect to $r$.

In order that cyclones should be possible the condition (\ref{v2})
should be fulfilled at least near the center of the vortex.
In the centers
\begin{equation}
V_r ~=~\tilde V_r ~=~ C_r \, \cos (2\varphi - F_r) ~=~ 0 \,.
\label{v3}
\end{equation}
This equation gives a possibility to reduce the left hand side of inequality
(\ref{v2}) to a form more suitable for an estimation.

In the vicinity of the center of the cyclone in residual velocities
(Fig.~2\,b) on the corotation circle from (\ref{12}) and (\ref{17}) we
have $F_\varphi$ $=$ $F_r+\pi$. Thus with the help of (\ref{v3}),
the condition (\ref{v2}) is reduced to
\begin{equation}
\left| F'_\varphi \, C_\varphi \right| ~>~ \left| V'_{circ} \right|
\,.
\label{v4}
\end{equation}
As follows from Fig.~5\,b at the corotation radius ($r$ $\approx$ $42$
arcsec) $C_\varphi$ $\approx$ $0$. Hence, as the value
of $F'_\varphi$  is very small the inequality (\ref{v4}) cannot be fulfilled.

In the case of "external" cyclones (see Fig.~2\,d) from (\ref{16aa}) and
(\ref{17}) we have $F_\varphi$ $=$ $F_r+\pi/2$. Then  using (\ref{v3})
from (\ref{v2})  we obtain
\begin{equation}
\left| C'_\varphi \right| ~>~ \left| V'_{circ} \right| \,.
\label{v4a}
\end{equation}
This relation explains the coincidence of the location of the external
cyclones in NGC~3631 (52 $<$ $r$ $<$ 59 arcsec in Fig.~7) with the region
of rapid growth of $C_\varphi$ (50 $<$ $r$ $<$ 60 arcsec in Fig.~5b).

In the case of the "internal" cyclones from (\ref{14}) and
(\ref{17}) it follows $F_\varphi$ $=$ $F_r-\pi/2$. Then in this case the
condition (\ref{v2}) also has the form (\ref{v4a}). According to Fig.~5b
it is fulfilled in the region 22 $<$ $r$ $<$ 31 arcsec, which
really contains the internal cyclone region 25 $<$ $r$ $<$ 30 arcsec
(Fig.~7).

The vector velocity field of the galaxy NGC~3631, when $V_{rot}(r)$
is taken in the form presented in Fig.~3 by the thickest line, is
shown in Figs.~6 and 7.
In the figures some streamlines of the velocity fields are presented.
If the velocity field is stationary the trajectories of fluid particles
coincide with the correspondent streamlines. One can see that the residual
velocity field of NGC~3631 (Fig.~6) corresponds to that of Fig.~2b and the
full velocity field in the reference frame of the spirals (Fig.~7)
corresponds to either of two cases presented in Fig.~2d when the centers
of two cyclones lie outside the corotation circle.  Internal cyclones in
the case shown in Fig.~7 are not enveloped by close streamlines.

\section{Conclusions}

I. An analysis of velocity fields in Grand-Design galaxies
shows that:

1) the field of residual velocities contains two cyclones and two
anticyclones with centers on the corotation circle;

2) in the reference frame rotating with the spiral pattern the
velocity field belongs to one of two types:

\noindent a) under the condition (\ref{1a}) it contains only two
anticyclones with centers near the corotation circle;

\noindent b) under the condition (\ref{1b}) besides the anticyclones
the field contains also either two or four cyclones. In the former case
the cyclone centers lay either on, or outside, or inside the corotation
circle. In the latter case, two pairs  of cyclones appear with centers
inside and outside the corotation circle.

II. Our analysis of the velocity field data obtained by our team at the 6m
telescope in the Special Astrophysical Observatory of the Russian Academy
of Sciences shows that the Grand-Design galaxy NGC 3631 belongs to the
type b).

\section{Acknowledgement}

We thank V.\,I.\,Arnold, B.\,V.\,Chirikov, V.\,L.\,Polyachenko,
M.\,I.\,Rabinovich, Ya.\,G.\,Sinai,  and specially G.\,Contopoulos for
fruitful discussions. We thank also J.Bolesteix for his kind placing
at our disposal a collection of interferometric filters.

This work was performed under partial financial support of RFBR grant
99--02--18432, grant "Leading Scientific Schools" 96--15--96648, and the
grant "Fundamental Space Researches. Astronomy" for the 1999 1.2.3.1 and
1.7.4.3.

\newpage
{\bf Figure Captions}

\vspace{5mm}
Figure 1. A scheme for the anticyclonic vortices formation in a
reference frame rotating with an arbitrary angular velocity.
Unperturbed rotation velocities and radial perturbed velocities are
marked by solid arrows and dashed arrows correspondingly. a) In the
laboratory system of coordinates the rotation velocity of the disc
varies with radius without change of sign. In this case the radial
perturbed velocity can not provide a vortex formation despite the
regular change of its sign along azimuth. b) In a rotating reference
frame, where the dotted circle is at rest, the same field of radial
perturbed velocities as in Fig.1a participates in the anticyclone
formation.

\vspace{5mm}
Figure 2. A scheme for the velocity field with vortices in different
reference frames.  Vectors of unperturbed rotation velocity field are
shown by solid arrows, radial and azimuthal components of
the residual velocity field --- by dashed arrows. Solid curves of different
thickness -- the thickest, the thinnest, and intermediate --
tracing the azimuthal locations of the maxima, minima and zero values of
the perturbed surface density defined at every radius are denoted by
$\tilde \sigma_{max}$, $\tilde \sigma_{min}$, and $0 (\tilde \sigma)$
respectively.  {\it A} and {\it C} denote anticyclones and cyclones
correspondingly. a) The angular velocities of a disc ($\Omega (r)$) and of
the spiral pattern ($\Omega_{ph}$) in the laboratory reference frame.
The dashed line represents the corotation circle. b) The residual velocity
field, which is the result of
the subtraction of the rotation velocity from the full velocity field.
Dotted lines show the boundaries of vortices.  c) The velocity
field in the reference frame rotating with the angular velocity
$\Omega_{ph}$ of the two-armed spiral pattern.  Only two anticyclones can
be seen in the vicinity of the corotation circle, when the gradient of
perturbed azimuthal velocity is lower than the rotation velocity gradient.
d) The velocity field in the same reference frame as in the previous
figure, but in the case, when the gradient of the perturbed azimuthal
velocity exceeds the rotation velocity gradient. In this case one can see
two cyclones and two anticyclones. Depending on the residual velocity
field geometry, the domination of the cyclonic shear over the anticyclonic
one can take place either near corotation, or inside (or outside) the
corotation circle.  Figure~d) demonstrates one more possibility when four
cyclones exist simultaneously - two inside and two outside the corotation
circle together with two anticyclones situated at corotation.

\vspace{5mm}
Figure 3. Examples of trial curves used to represent the rotation curve
($V_{rot}(r)$) in NGC~3631 are shown by solid lines together with the observed
behaviour of $a_1^{obs}(r)$ (triangles). The thickest line marks the
rotation curve corresponding to a quasi-stationary gaseous disk.

\vspace{5mm}
Figure 4. The radial dependence of $b_3^{obs}-b_1^{obs}$ $=$ $C_r
\, \cos \,F_r$ and $b_3^{obs}+b_1^{obs}$ $=$ $C_\varphi \, \sin
\,F_\varphi$ observed in the spiral galaxy NGC~3631. An estimation of
the amplitudes of the velocity components from the extrema of the presented
functions gives $max(C_r)$ $\simeq$ $max(C_\varphi)$ $\simeq$ 60~km/s.

\vspace{5mm}
Figure 5. Some parameters of the  velocity field which correspond to a
quasi-stationary regime of the vortex structure.

(a) The calculated amplitude of the radial velocity as a function of $r$.

(b) The same for the azimuthal velocity (squares) with overlaid profile of
$\left| V_{circ}(r)\right|$.

\vspace{5mm}
Figure 6. The residual velocity field in the plane of the gaseous disc of
the galaxy NGC 3631. One can see two cyclones and two anticyclones.
Solid lines show some streamlines of the residual velocity
field.

\vspace{5mm}

Figure 7. The full velocity field of the same galaxy in the reference
frame rotating with the angular velocity of the two-armed spiral pattern.
Overlayed squares show the position of maxima of the second Fourier
harmonics of brightness map of NGC~3631 in H$_\alpha$ line. Thin circle
shows the position of the corotation. Presented streamlines (solid curves)
were calculated using reverse "time" direction that allows to
reveal separatrices evidently. Two cyclones and two anticyclones are
enveloped by close streamlines. Other two cyclones are not enveloped by
close streamlines. The anticyclone centers lie almost at corotation;
the cyclone centers lie outside corotation  Cyclones lie in the vicinity
of spiral arms, anticyclones are situated between them. The agreement of
the vortices position with theoretical predictions (scheme in Fig.~2d) is
very good.


\newpage

\begin{thebibliography}{99}

\bibitem{1} V.\,L.\,Afanasiev and A.\,M.\,Fridman, Astron. Lett. 19 (1993)
319.

\bibitem{2} A.\,M.\,Fridman, O.\,V.\,Khoruzhii, V.\,V.\,Lyakhovich,
V.\,S.\,Avedisova, O.\,K.\,Sil'chenko, A.\,V.\,Zasov, A.\,S.\,Rastorguev,
V.\,L.\,Afanasiev, S.\,N.\,Dodonov and J.\,Boulesteix, Astrophys. Space Sci.
252 (1997) 115.

\bibitem{3} G.\,Contopoulos, Astrophys. J. 181 (1973) 657.

\bibitem{4} P.\,V.\,Baev, Yu.\,N.\,Makov and A.\,M.\,Fridman, Sov. Astron.
Lett.  13 (1987) 406.

\bibitem{5} G.\,Bertin and C.\,C.\,Lin, Spiral structure in galaxies: a
density wave theory (The MIT Press; Cambridge, Massachusetts; London,
England, 1995).

\bibitem{6} A.\,M.\,Fridman, O.\,V.\,Khoruzhii, A.\,V.\,Zasov,
O.\,K.\,Sil'chenko, A.\,V.\,Moiseev, A.\,N.\,Burlak,
V.\,L.\,Afanasiev, S.\,N.\,Dodonov and J.\,Knapen, Astron. Lett 24 (1998)
764.

\bibitem{6aa}  V.\,V.\,Lyakhovich, A\,.M.\,Fridman,  O.\,V.\,Khoruzhii,
Astron. Rep. 40 (1996) 18.

\bibitem{6a} A.\,M.\,Fridman, O.\,V.\,Khoruzhii,
V.\,V.\,Lyakhovich, O.\,K.\,Sil'chenko, A.\,V.\,Zasov, V.\,L.\,Afanasiev,
S.\,N.\,Dodonov and J.\,Boulesteix, Astron. \& Astroph. (submitted).

\bibitem{7} V.\,V.\,Lyakhovich, A.\,M.\,Fridman, O.\,V.\,Khoruzhii and
A.\,I.\,Pavlov, Astron. Rep. 41 (1997) 447.

\bibitem{8} B.\,Lindblad and R.\,G.\,Langebartel. Stockholms Observatoriums
Annaler, Band 17 (1953) 6.


\bibitem{10} A.\,Fridman, ASP Conference Series, 66 (1994) 15.

\bibitem{11} D.\,Lynden-Bell, Barred galaxies and circumnuclear activity:
proceedings of the Nobel Symposium 98, eds. A.\,Sandquist \&
P.\,O.\,Lindblad (Springer-Verlag, Berlin, Heidelberg, New-York, 1996).

\bibitem{cont} G.\,Contopoulos, J.\,Hunter, and M.\,England, to be
published.

\bibitem{9} A.\,M.\,Fridman, O.\,V.\,Khoruzhii, E.\,V.\,Polyachenko,
A.\,V.\,Zasov, O.\,K.\,Sil'chenko, A.\,V.\,Moiseev, A.\,N.\,Burlak,
V.\,L.\,Afanasiev, S.\,N.\,Dodonov and J.\,Knapen, to be published.

\bibitem{12} A.\,A.\,Sumin, A.\,M.\,Fridman, and V.\,A.\,Haud, Pis'ma
Astron. Zh. 17 (1991) 698.
\end{thebibliography}
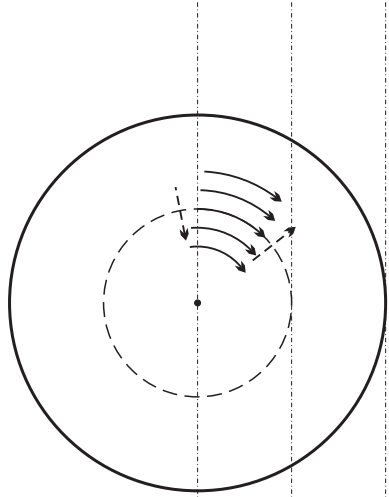
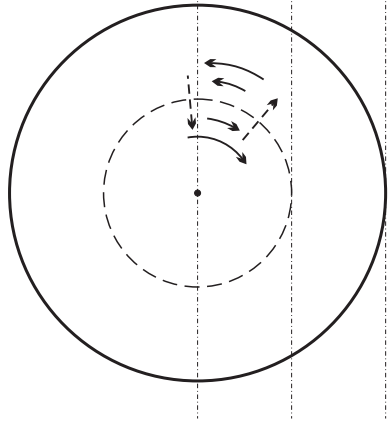
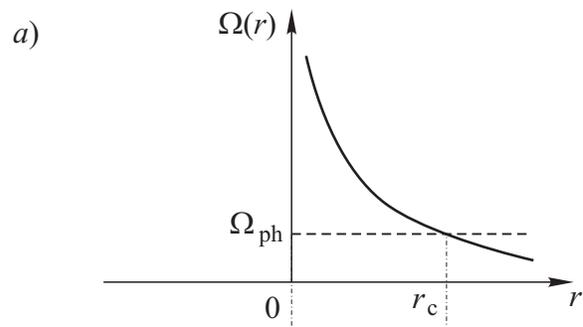
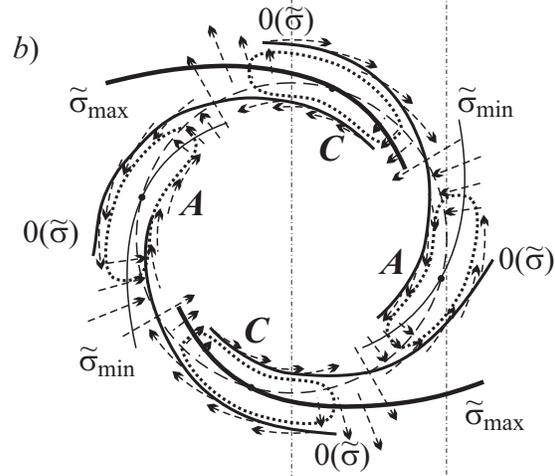
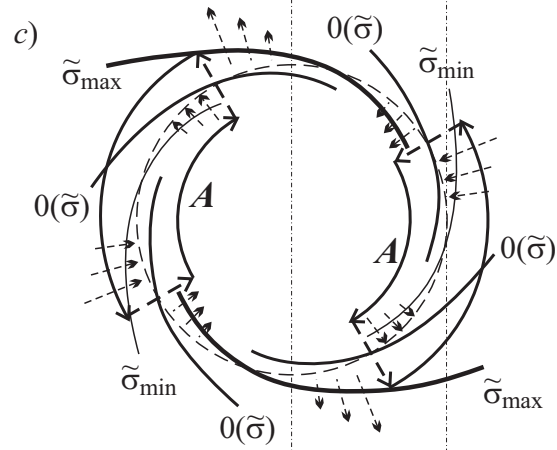
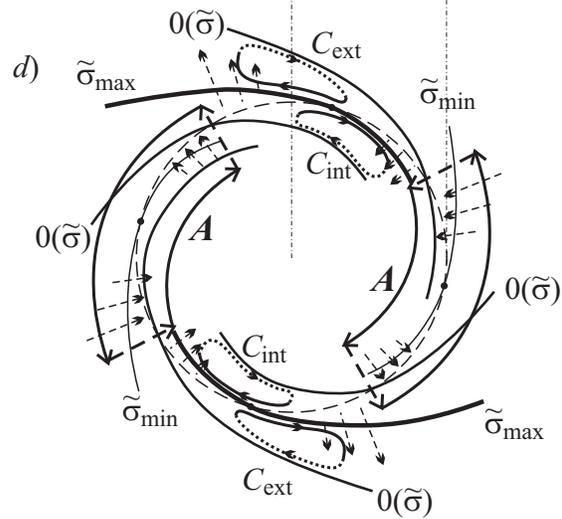
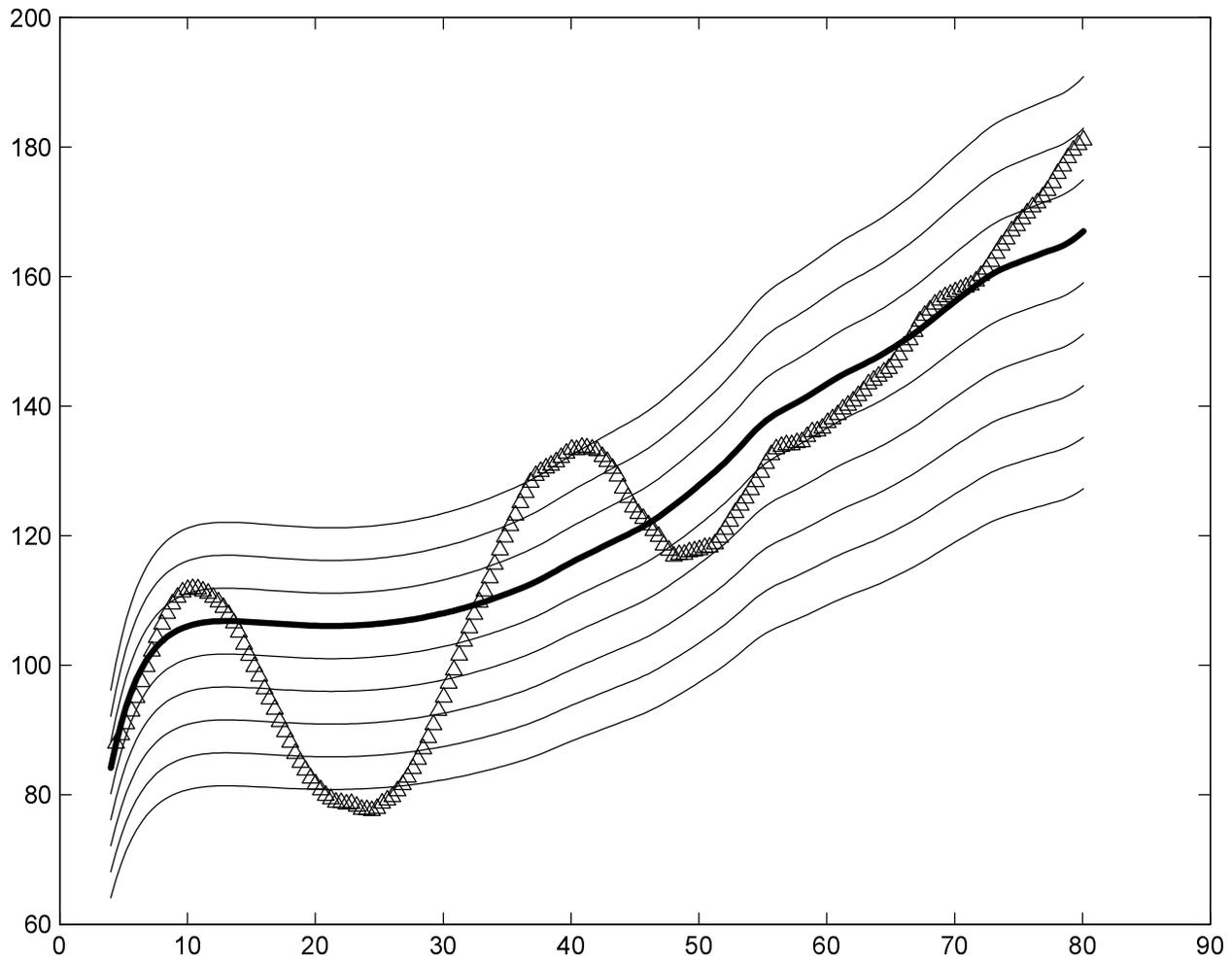
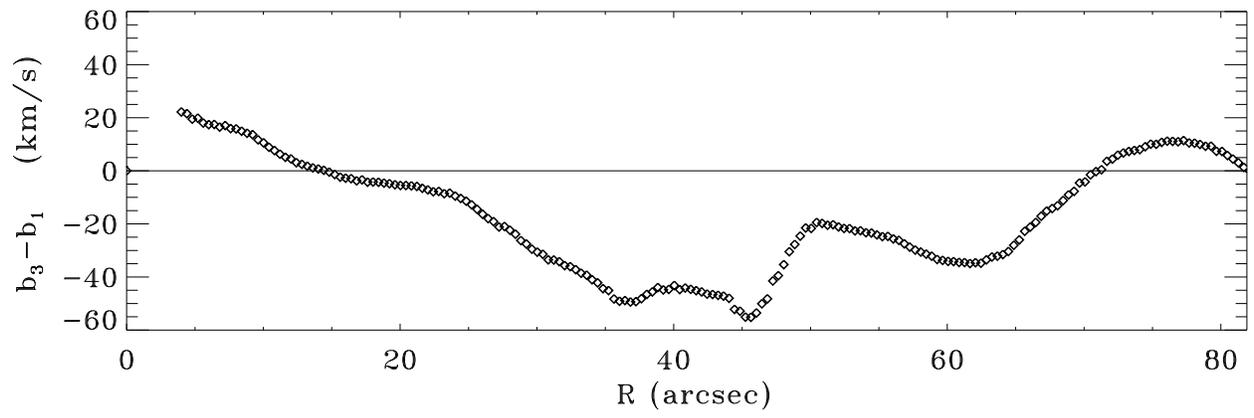
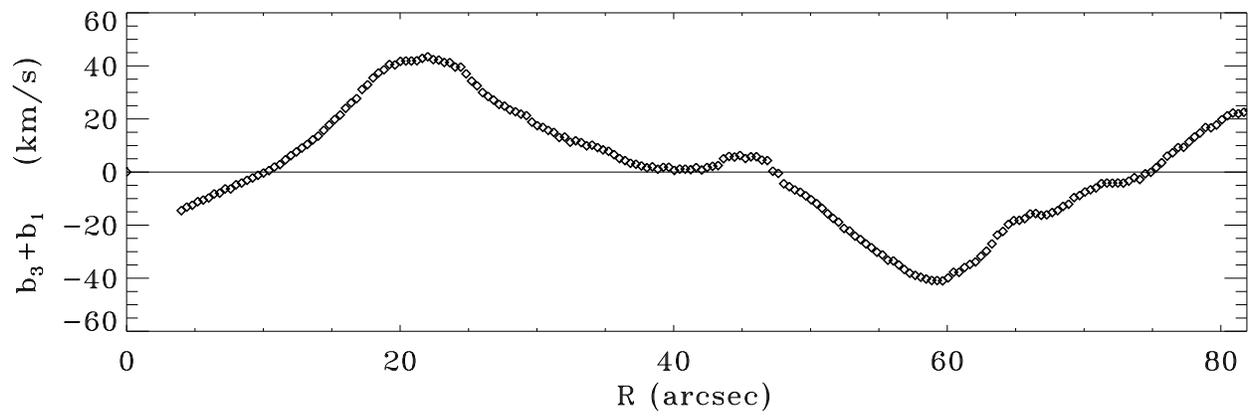
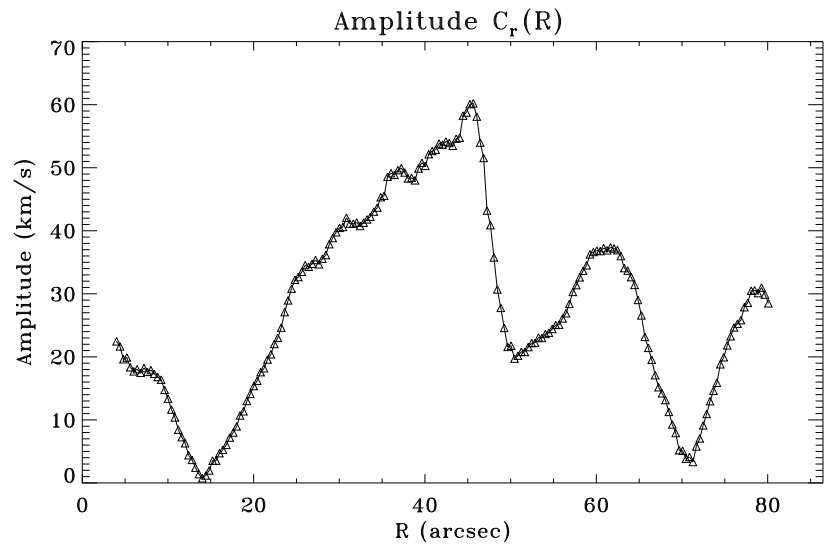
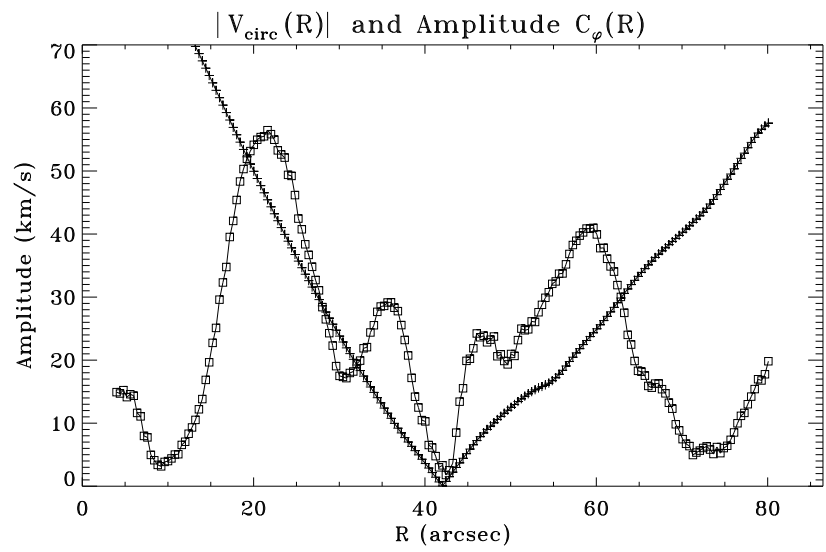
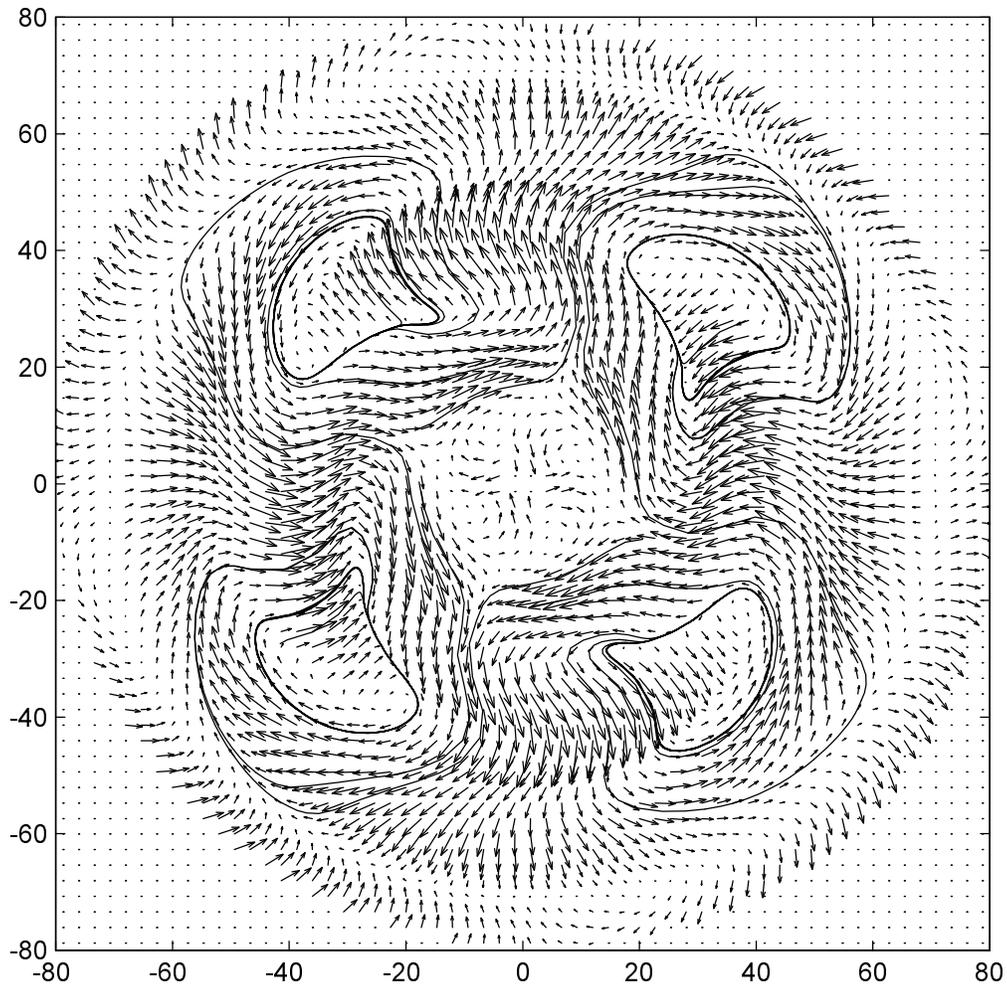
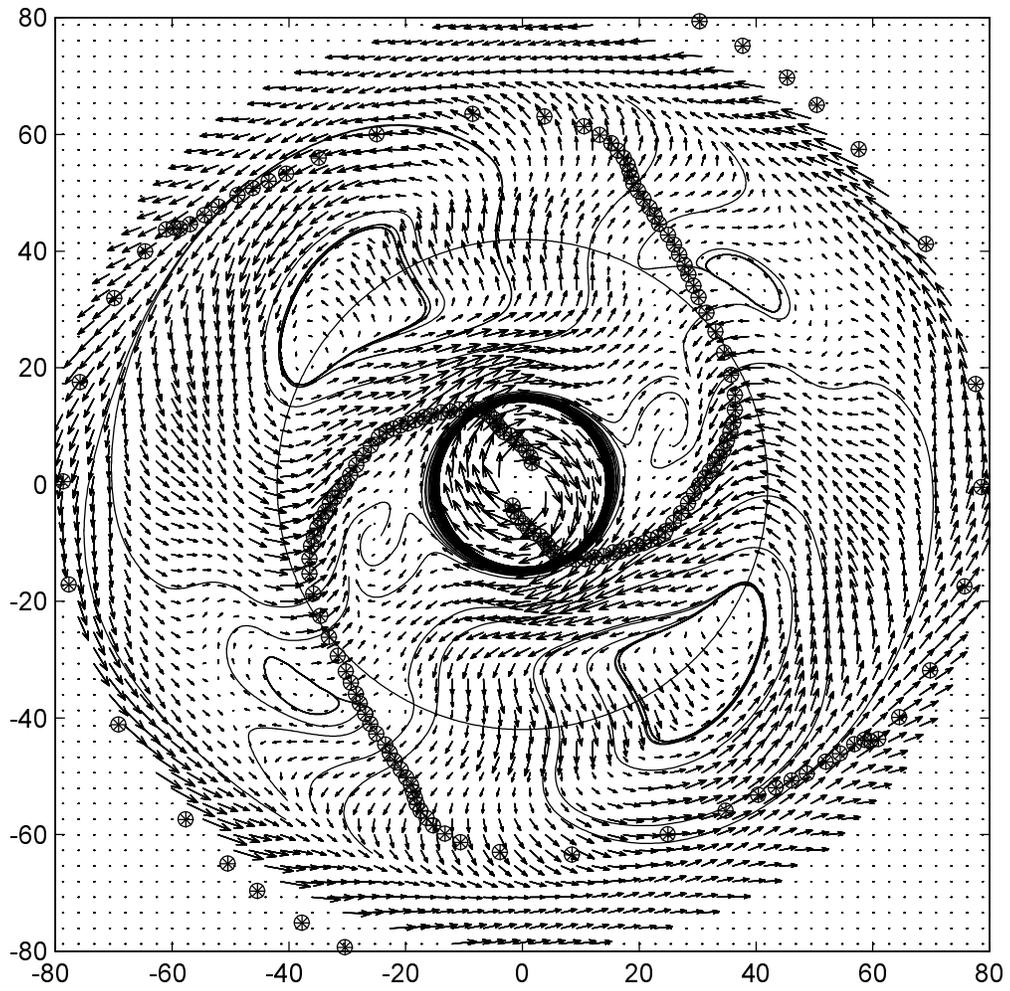
\end{document}